# Electrically Pumped Single Transverse-Mode Coupled Waveguide Laser by Parity-time (PT) Symmetry

Ruizhe Yao, Chi-Sen Lee, Viktor Podolskiy, and Wei Guo*

Physics and Applied Physics Department, University of Massachusetts Lowell, Lowell, MA, 01854, USA.

*Correspondence to: wei_guo@uml.edu

**Abstract**: We demonstrate single transverse-mode operation of InAs quantum dot (QD) broad-area coupled waveguide lasers enabled by parity-time (PT) symmetric breaking. A novel electrically pumped PT laser operates on waveguide cavity with parallel gain and loss regions. Such counterintuitive waveguide design enables PT symmetric breaking, causing unique mode selection and ultimately enabling single mode operation. By tuning the loss in the loss region of the coupled waveguide cavity, several different PT-symmetric regimes are analyzed theoretically, and the corresponding PT symmetric phase transition is demonstrated experimentally in the coupled waveguide laser.

In recent explorations, parity-time (PT) symmetry offers a unique pathway to advance laser science by incorporating counterintuitive gain and loss across the laser cavity (*1-10*). Several promising experimental demonstrations of PT-symmetry-based lasers have been reported (*11-18*). For example, Zhang et. al. (*19*) and Khajavikhan et. al. (*20*) have shown optically pumped single mode microcavity and micro-ring lasers operated in the PT symmetric broken regime, respectively, where the gain is achieved by optically pumping the active regions and loss is obtained from material absorption loss. Unfortunately, up until now, applications of PT symmetry in lasers are mostly limited to optically injected devices (*21, 22*). Such devices lack effective ways to dynamically adjust gain and loss in the cavities, which is necessary to reveal and fully utilize the rich physics of PT symmetry in optics. In this context, here we report a coupled waveguide PT symmetric laser with independently and electrically tuned gain and loss. The PT symmetric mode discrimination conditions are analyzed numerically and confirmed experimentally. Several distinct phases of PT symmetry breaking are identified and the corresponding cavity mode manipulation by PT symmetry is observed experimentally. Most importantly, we demonstrate an electrically injected single transverse-mode broad-area coupled waveguide lasers enabled by PT symmetric mode selection.

In the applications of high power lasers and tapered amplifier (TA) diodes, it is essential to utilize large-area waveguides while still maintaining single transverse-mode operations (*23*). However, present design paradigms are accompanied by degraded laser performance, unstable single mode operation at high output power and extra fabrication complexity due to the tapered waveguide shapes (*24, 25*). In this context, PT symmetry provides a unique pathway to manipulate the laser cavity modes and realize a single transverse-mode operation without sacrifice of the laser performance.

Laser cavity used in this work comprises two electrically isolated quantum dot (QD)-filled Fabry-Perot broad-area laser cavities. The two regions can be considered as two coupled (multimode) waveguides, due to the large width. The concept of PT-symmetry-breaking mode manipulations in coupled waveguides was first theoretically discussed by Miri et. al. (*26*). In the coupled waveguide cavities, a broad-area waveguide with gain $g_m$ is coupled to its counterpart of loss $\alpha_m$ with coupling coefficient $\kappa_m$ It has been shown that the combination of $g_m/\alpha_m$ and $\kappa_m$ determines the PT symmetry breaking conditions. When $g_m < \kappa_m$, the modes of the coupled waveguide represent symmetric and anti-symmetric combination of the modes of the two waveguide components. The propagation constants of the two combinations are often degenerate featuring complete balance of gain and loss. However, when $g_m > \kappa_m$, the PT symmetry is spontaneously broken, and modes of the waveguide components become the modes of the combined waveguide. Most importantly, only one of the two supermodes exhibits gain. In Fabry-Perot cavities, the higher order modes are typically of higher coupling coefficient than the one of the fundamental mode. Therefore, it is possible to design a coupled waveguide cavity to allow only the fundamental mode to reach the PT symmetry breaking threshold, hence, only the fundamental mode can have net gain and lase.

Although ideal PT symmetric cavities require equal gain and loss, semiconductor lasers generally exhibit gain clamping after lasing, a process that limits cavity gain at the threshold modal gain, $g_{th}$ (*26, 27*). Therefore, it is essential to explore PT symmetry in coupled waveguides with fixed gain and varied loss configurations.

Here, the dependence of mode profiles and their eigenfrequency was analyzed with commercial finite-element-method solver of Maxwell equations, COMSOL Multiphysics®. Figure 1a illustrates the dependence of imaginary part of the eigenfrequency as a function of modulation of loss in one half of the waveguide, while the gain of 4.95 $cm^{-1}$ is fixed in the other half. Similar to the idealized PT-symmetry breaking studies (*28*), a series of the exceptional points (EPs) can be identified.

As shown in Figure 1a, the coupled waveguide PT symmetric laser exhibits several phase transitions. The corresponding mode electric field distribution in each phase is shown in Figure 1b-1e. In Phase I, $\alpha < 1.8$ $cm^{-1}$, all the modes are below their EPs. Due to the asymmetric gain and loss, the non-broken modes exhibit net gain in this phase, and the coupled waveguide cavity behaves as a conventional broad-area cavity supporting multiple transverse modes to lase. When the loss is between 1.8 and 4.95 $cm^{-1}$ (Phase II), the first pair of supermodes have reached their EP, but, since the total gain is still greater than loss, the broken supermodes as well as some non-broken modes show net gain, and single mode operation is not yet obtained. When the loss is further increased from 4.95 $cm^{-1}$ to 16.9 $cm^{-1}$, only one pair of the supermodes shows net gain, making the single mode operation possible, as indicated in Phase III. When the loss reaches 16.9 $cm^{-1}$, Phase IV, the second pair of supermodes pass their EP. However, since the gain is not large enough to overcome loss in this phase, single mode operation is still preserved. Finally, with further increment of the loss approaching to 44 $cm^{-1}$, one of the second pair of the PT symmetry broken supermodes have sufficient gain to overcome loss, and single-mode operation is prohibit again, as shown in Phase V.

The coupled waveguide PT symmetric laser is experimentally demonstrated as well. Shown in Figure 2a, the QD laser heterostructures is grown by molecular beam epitaxy (MBE). Figure 2b and 2c show the schematic and oblique view of the coupled waveguide laser, respectively. The coupled waveguides with total width of 60 μm are obtained and two p-type Ohmic contacts are defined on top of the coupled waveguides to provide independent control of gain and loss in the waveguides.

The near- and far- field patterns of the coupled waveguide lasers are measured to characterize the cavity modes and laser output beam profile. Figure 3a-3c show the near- and far-field patterns of the laser diodes with the gain bias current of 400 mA and loss bias current at 0, 50 and 120 mA, respectively. It is worth noting that $I_{th}(L=\infty)$ of the coupled waveguide lasers of 132 mA is fitted from independent experiments with the same waveguide geometry. Thus, although the loss waveguide is under small forward bias, it still exhibits loss in our cases. In the far-field pattern in Figure 3a, it is clearly seen that, with zero biased loss waveguide, the coupled waveguide laser exhibits clear signs of high order modes as indicated by the label A and B, where the fundamental mode is labeled as C. The near-field pattern shows that the emission is from the gain waveguide facet. It is estimated that the total waveguide loss in the loss cavity is > 50 $cm^{-1}$ by considering the InAs QD absorption loss under zero bias (*29*). Thus, due to the large loss and small gain, the coupled waveguide laser is operated in Phase V. When the bias current of the loss waveguide is increased to 50 mA, the peaks from the higher order modes are successfully suppressed as shown in Figure 3b and near single-lobe far-field pattern is obtained in the broad-area coupled waveguide lasers. This indicates that the laser is now operating in Phase III or IV. In addition, the peak output power is reduced by 17 %, which is due to the suppression of higher order modes. Finally, when the loss waveguide bias is further increased to 120 mA, higher order mode lobes reappear in the far-field pattern, Figure 3c. The far-field pattern exhibits small shift towards the center of the coupled waveguide, and multiple lobes in the pattern imply that higher order un-broken modes start to lase compared to the case in Figure 3a. The near-field pattern shows that the laser emission pattern starts to shift towards the center of the coupled waveguides and a broken fundamental mode is still observed in the gain waveguide. This indicates that the device is now operating in Phase I/II, where both the broken fundamental mode and unbroken higher order modes lase and the former has higher power due to larger gain compared to the rest lasing modes. It is worth noting that the higher order modes shown in Figure 3a is less profound than the ones in Figure 3c, this is due to that the gain of higher order modes are naturally suppressed and determined by the loss in the lossy waveguide in the PT symmetry broken phase. This is fundamentally different from the traditional tapered waveguide lasers, where the gain of higher order modes can still increase and exhibit lasing at higher bias conditions. Thus, it is anticipated that, if the proposed coupled waveguide PT symmetric lasers are further optimized, e.g. more proper thermal management, the PT symmetric coupled waveguide lasers can exhibit higher output power limits.

In summary, in this work, we have experimentally demonstrated a single transverse-mode broad-area coupled waveguide laser based on PT symmetry. By changing the loss in the waveguide, the supermode broken and phase transitions are experimentally observed and agree well with the theoretically predication. Further work is under way to optimize the device efficiencies and reduce the device Joule heating.

Nevertheless, this work opens a pathway for the practical device applications of PT symmetry in optics.

**Figure Caption**

**Figure 1:** (a). Imaginary part of the eigenfrequency vs. loss in the loss waveguide. Inset: Enlarged region of $\alpha < 4.95$ $cm^{-1}$. The red dashed lines indicate zero propagation constant. In Phase I, all the modes are below their EPs. In Phase II, the first pair of supermodes have already reach their EP, but, since the total gain is still greater than loss, the broken supermodes as well as some non-broken modes have net gain. In Phase III, only one pair of the supermodes have net gain and the single mode operation becomes possible. In Phase IV, the second pair of supermodes pass their EP. However, since the gain is not large enough to overcome loss in this phase, single mode operation is still preserved. In Phase V, one of the second pair of the PT symmetry broken supermodes has sufficient gain to overcome loss and single mode operation becomes prohibit again. (b)-(e): Electric field distribution of $TE_0$ and $TE_1$ mode with the loss of 1 $cm^{-1}$, 10 $cm^{-1}$ and 30 $cm^{-1}$ in the loss waveguide.

**Figure 2:** Heterostructures (a) and schematic (b) of the InAs QD PT symmetric coupled waveguide laser; (c) SEM overview of the fabricated InAs QD coupled waveguide laser.

**Figure 3:** (a)-(c) Far- and near- (inset) field patterns of coupled waveguide laser with $I_{gain}$= 400 mA, and $I_{loss}$ varying from 0 to 120 mA; The fundamental mode (C) and higher order modes (A and B) are labelled. It is seen that, with zero biased loss waveguide, the coupled

waveguide laser exhibits clear signs of high order modes as indicated by the label A and B. The near-field pattern shows that the emission is from the gain waveguide facet. The coupled waveguide laser is operated in Phase V where both the fundamental and first order modes have pass their PT symmetry EP. When the bias current of the loss waveguide is increased to 50 mA, the peaks from the high order modes are successfully suppressed and near single-lobe far-field pattern is obtained in the broad-area coupled waveguide lasers. This indicates that the laser is now operating in Phase III/IV. When the loss waveguide bias is further increased to 120 mA, high order mode peaks reappear in the far-field pattern. The far-field pattern exhibits small shift towards the center of the coupled waveguide, and the multiple peaks imply that high order un-broken modes start to lase. This implies that the laser is now operating in the Phase I/II.

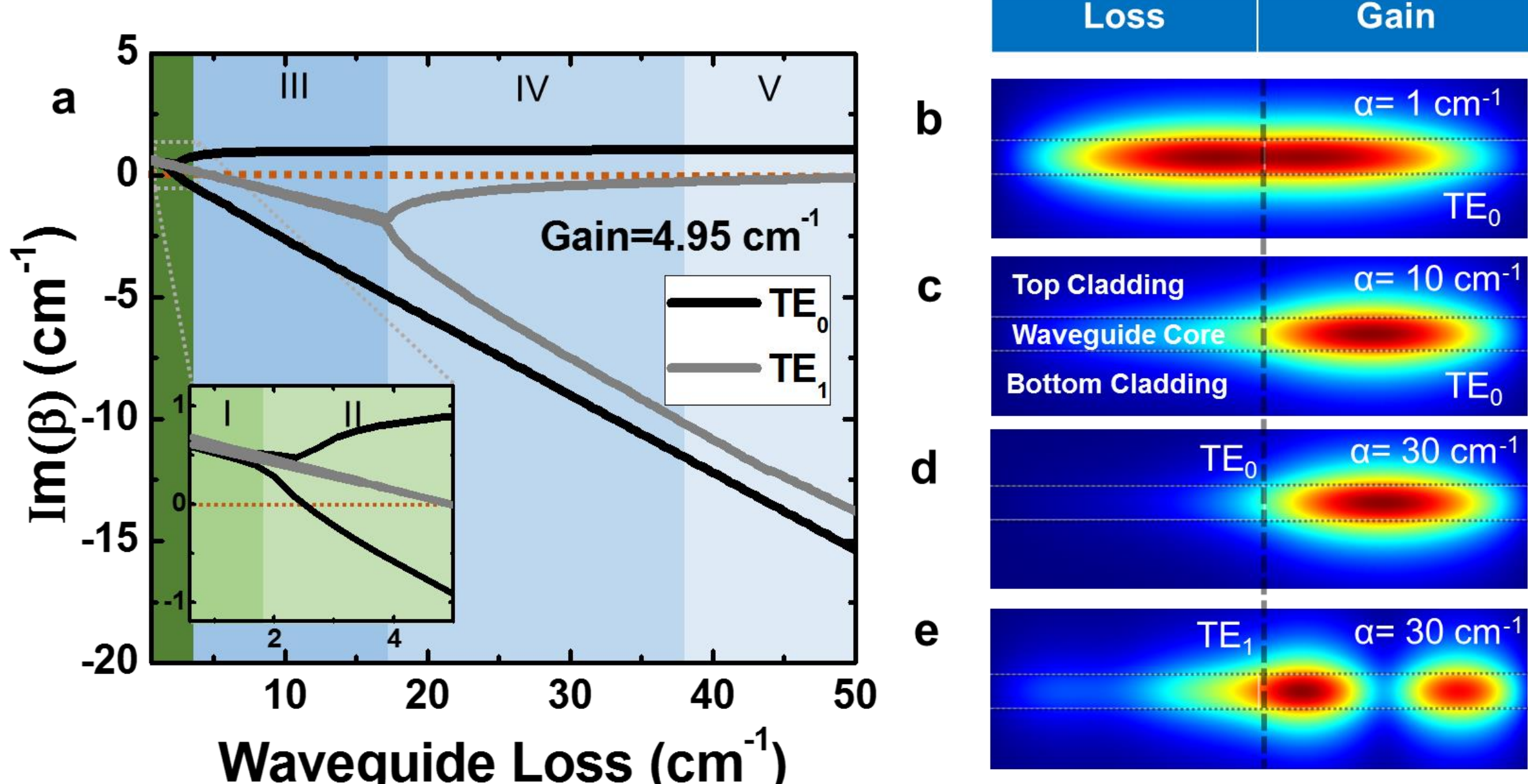

Figure 1 of Yao *et. al.*

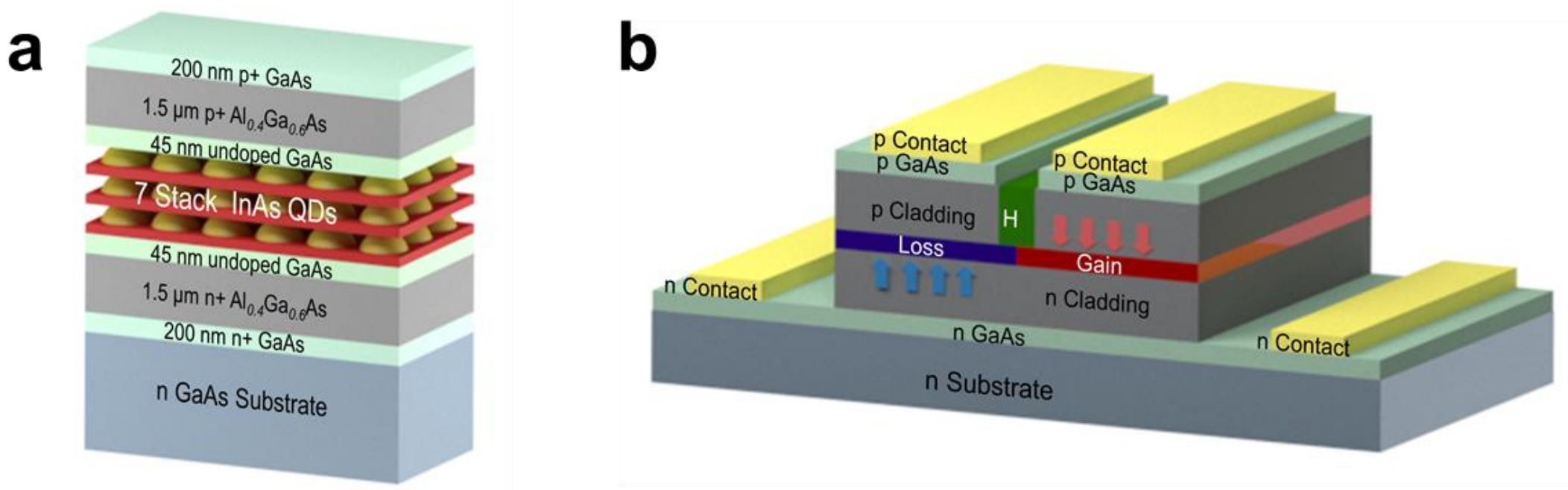

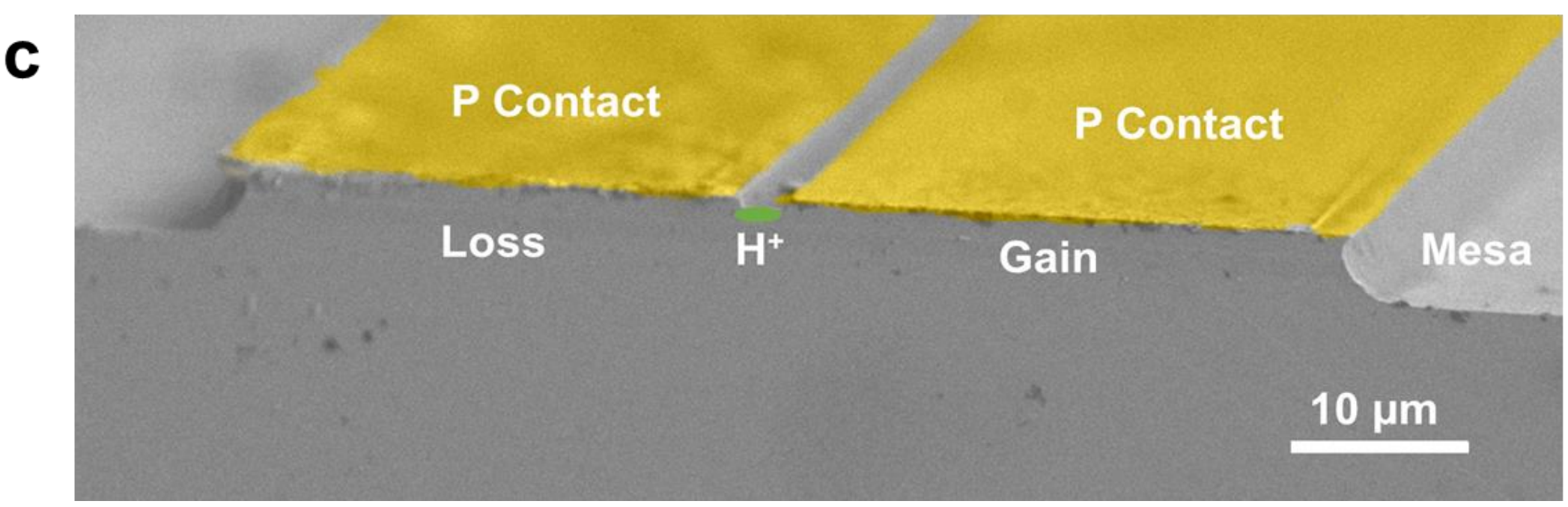

Figure 2 of Yao *et. al*

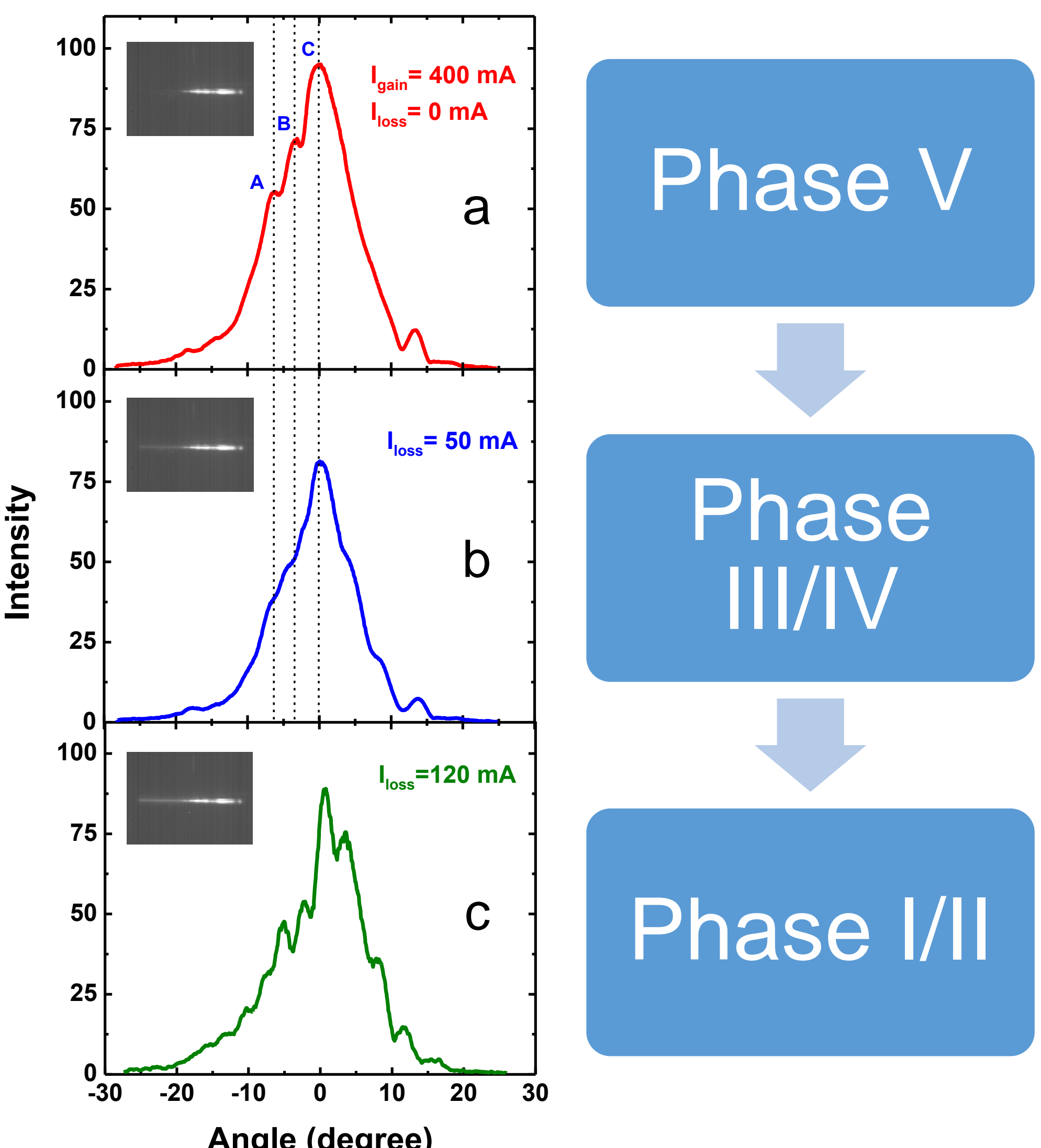

Figure 3 of Yao *et. al.*